\documentstyle[11pt,twoside]{article}
\begin{document}

\noindent Accepted for publication in the Proceedings of the Astronomical
Society of Australia,
vol 12, 1995.

\begin{center}
POSSIBLE CONNEXION BETWEEN LARGE SCALE STRUCTURES\\

H. DI NELLA$^{1,2}$ and G. PATUREL$^{1}$\\

Observatoire de Lyon$^{1}$, F69230 Saint-Genis Laval, France\\
School of Physics$^{2}$, Univ. of New South Wales, Sydney 2052, Australia\\

\end{center}

\input epsf

\begin{center}
ABSTRACT
\end{center}
 From the Lyon-Meudon Extragalactic Database LEDA, we constructed a sample
of 27,000 galaxies with radial velocities smaller than 15,000 km s$^{-1}$. From
this first sample we selected the 5,683 largest galaxies which constitute
a sample complete up to the diameter limit $D_{25}=1.6'$. This sub-sample
has been used to search for the most populated plane. The result is that
the plane defined by the pole ($l=52 \deg$; $b=16 \deg$) contains twice
as many galaxies as would be found for randomly distributed galaxies.
This plane is not far from the supergalactic plane.

Then, the distribution of galaxies is studied in this plane by using the
radial velocity as an estimate of the distance. An excess of galaxies
appears at a distance of about 70 Mpc. We discuss the reality of this
density enhancement which could result from selection effects.\\

Subject headings: galaxies: redshifts - large scale structure of the universe

\section{Introduction}
Our long-term goal is a study of the kinematics of the local
Universe (z$<$0.05). Such a study is obviously connected with the
distribution of galaxies around our Local Super Cluster (LSC). In order
to make it easier we searched first for the most populated plane to reduce
the study to a 2D case. This is suggested by preliminary results
(Bottinelli et al. 1986; Tully 1986) showing that a dominant
concentration of galaxies is located in a plane close to the supergalactic
one.

The first step is thus the construction of the sample. It is particularly
important to get a sample with a well-defined completeness limit in order
to study selection effects. For this purpose we used the Lyon-Meudon
Extragalactic Database LEDA
\footnote {To connect LEDA enter {\it telnet lmc.univ-lyon1.fr}
login: {\it leda}}.
LEDA is a free access database which has been created in 1983 to collect
the main parameters of the principal galaxies.
More than 90,000 galaxies are managed with their cross-identifications and
observational measurements, including radial velocities and apparent
diameters which are the relevant parameters for the present study.

\section{The sample}
{}From LEDA, we thus constructed a sample of 27,000 galaxies with
radial velocities smaller than 15,000 km s$^{-1}$. From this first sample we
selected the 5,683 largest galaxies (containing all morphological types) which
actually constitute a complete
sample up to the diameter limit $D_{25}=1.6'$. The classical test of
completeness $\log N$ vs. $\log D_{25}$ (see Paturel et al.
1994) was performed from which we
conclude that the sample is complete at a level of 85-90\%.
It means that 85\% of galaxies
satisfying both conditions V$<$15000 km s$^{-1}$ and $D_{25}=1.6'$ are
included in the sample (it does not mean that  the sample
contains 85\% of all existing galaxies within the limiting distance
- an impossible task).
The major interest in having a sample with such a clear definition is that
it makes it possible to predict how many galaxies are expected in a given
direction assuming an isotropic distribution, or, how many galaxies are
expected at a given distance assuming a homogeneous distribution.
The comparison with the actual number will thus allows us to detect
possible difference with an homogeneous and isotropic distribution.

\section{The most populated plane }

Altering the orientation of a test plane in steps of $1 \deg$ in
galactic longitude and latitude, we
counted galaxies lying within $15 \deg$ of it. We found that
the plane defined by the pole ($l=52 \deg$; $b=16 \deg$), in galactic
coordinates, is the most populated one. More precisely, from the
covered solid angle
one should expect 25\% of the total
sample within the $\pm 15 \deg$ limits defining the plane (if one assumes
an isotropic distribution of galaxies), while the actual
percentage reaches 45\%. The conclusion is that the corresponding plane
contains most of the local visible masses. It contains the main clusters
and superclusters: Perseus-Pisces, Pavo-Indus, Centaurus, Coma and LSC.
This plane is close to the supergalactic plane or hypergalactic plane
(Paturel et al. 1988). The trace of this plane is shown
in a Flamsteed equal area projection (Fig. 1).

The conclusion of this section is that, within the considered volume, the
distribution is not isotropic but well concentrated in a dominant
plane which will constitute our new reference framework (hereafter,
MPP).

\section{Distribution of galaxies within the MPP}
For each galaxy of the complete sample we define the distance from
the heliocentric radial velocity, adopting a Hubble constant of
H=75 km s$^{-1}$ Mpc$^{-1}$. Then, the 5,683 galaxies are plotted in the
MPP. The result is shown in Fig. 2.
Apart from radial structures (known as "fingers of God") which
are obvious artefacts due to the velocity dipersion in clusters, a
cocoon-like structure can be seen around the LSC (Di Nella \& Paturel 1994).
Some parts of it can
be recognized as traces of already known
structures: e.g. The ''Great Wall'' (Geller \& Huchra 1988), The
''Southern Wall'' (or Pavo-Indus supercluster) (Da Costa et al.
1988) and the Perseus-Pisces wall
(Haynes \& Giovanelli, 1986).

In order to conduct a more quantitative analysis we made a histogram of
distances of our galaxy sample and compared it with the one predicted
from the selection function of a homogeneously distributed sample having
the same diameter limit. The result is shown in Fig. 3. A
very significant secondary peak appears around $\approx$ 68 Mpc. The
excess of galaxies seen in Fig. 2 is confirmed, but this
does not confirm that the cocoon shape is real.

The conclusion of this section is that the distribution of galaxies
within the hypergalactic plane does not look homogeneous. Is it real
or not? Is it just the result of selection effects?
This will be discussed in the last section.

\section{Discussion}
Several arguments can be given against the reality of the ring-like
structure seen in Fig. 2:
\begin{itemize}
\item The cocoon is almost centered on our galaxy. This anthropocentric
position cannot be accepted.
\item The density enhancement in Fig. 3 could be due to some
selection effects (e.g. the limiting diameter can be different for
field galaxies and cluster galaxies).
\item None of the other team working on the Large Scale distribution of
galaxies gives similar description.
\item The density enhancement of the LSC does not appear on IRAS map
suggesting that the LSC does not play a central role as depicted by
Fig. 2.
\end{itemize}

The corresponding counter-arguments are as follows:
\begin{itemize}
\item It is not easy to find the center of the cocoon. At least the
curved trace of the Great Wall is not centered on our Galaxy.
We cannot exclude that the center is in fact on the LSC.
\item The density enhancement in Fig. 3 comes from
some well established structures which appear to be connected by the
cocoon-like structure. We do not suspect that these well-admited
structures result from selection effects.
\item The cocoon-like structure can only be found when working with
a whole sky sample. It is also most easily visible by viewing a projection of
the most populated plane.
\item The LSC does exist.
\end{itemize}

In conclusion, even if the distribution of nearby galaxies looks like
an artefact, it is not clear what would be the cause of it. \\

We are grateful to those who take part in the management
of the LEDA extragalactic database: N. Durand, A.M. Garcia, R. Garnier,
M.C. Marthinet, C. Petit. We thank Don Mathewson, G\'erard de Vaucouleurs
and Warrick Couch
for their comments and discussions.

\begin{center}
REFERENCES
\end{center}

\noindent Bottinelli, L., Fouqu\'e, P., Gouguenheim, L., Paturel, G. 1986,
La Dynamique des Structures Gravitationnelles (Observatoire de Lyon)

\noindent Da Costa, N., Pellegrini, P.S., Sargent, W.L.W., Tonry, J.,Davis, M.,
Meiksin, A., Latham, D.W., Menziers, J.W., Coulson, I.A. 1988, ApJ 327, 544

\noindent Di Nella, H., Paturel, G. 1994, C.R.Acad.Sc. Paris, t.319, p57

\noindent Geller, M., Huchra, J. 1988, Science 246, 897

\noindent Paturel, G., Bottinelli, L., Gouguenheim, L., Fouqu\'e, P. 1988,
A\&A 189, 1

\noindent Paturel, G., Bottinelli, L., Di Nella, H., Fouqu\'e, P., Gouguenheim,
L.,
Teerikorpi, P. 1994, A\&A 289, 711

\noindent Haynes, M., Giovanelli, R. 1986, ApJ 306, L55

\noindent Tully, R.B., 1986, ApJ 303, 25

\newpage

\begin{figure}
\vspace{1.6cm}
\caption{Flamsteed equal area projection for 27,045 galaxies in supergalactic
coordinates. A S-shape shows the trace of the
most populated plane: the hypergalactic plane. This plane
connects Perseus-Pisces,
Pavo-Indus,  Centaurus, Coma and Virgo superclusters.}
\label{flam}
\end{figure}

\begin{figure}
\vspace{1.6cm}
\caption{Distribution of the Leda galaxies in
the most populated plane.}
\label{planxy}
\end{figure}

\begin{figure}
\vspace{1.6cm}
\caption{Histogram of distances around the LSC. A homogeneously
distributed sample would have given the dashed line curve.}
\label{histo}
\end{figure}





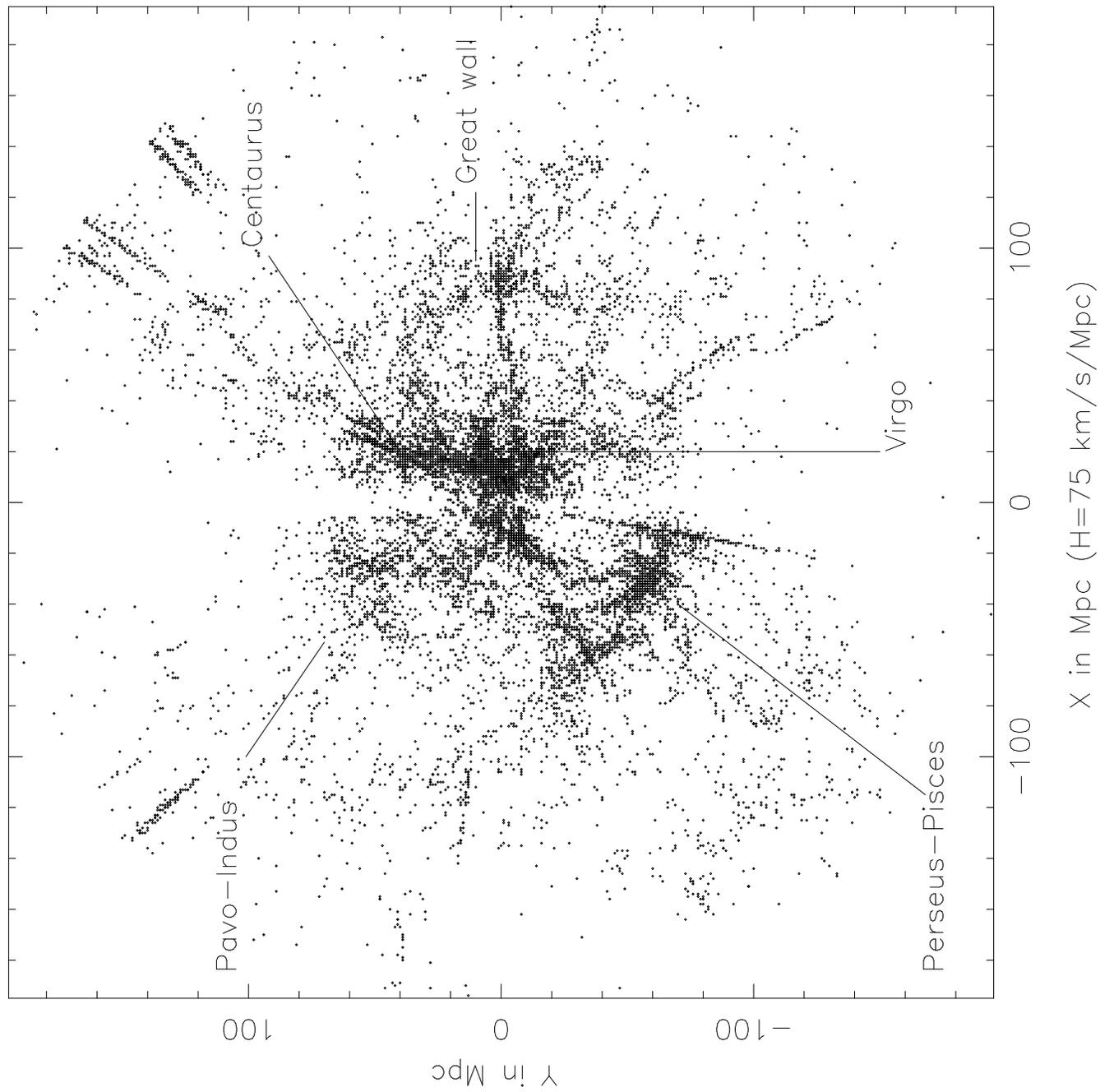
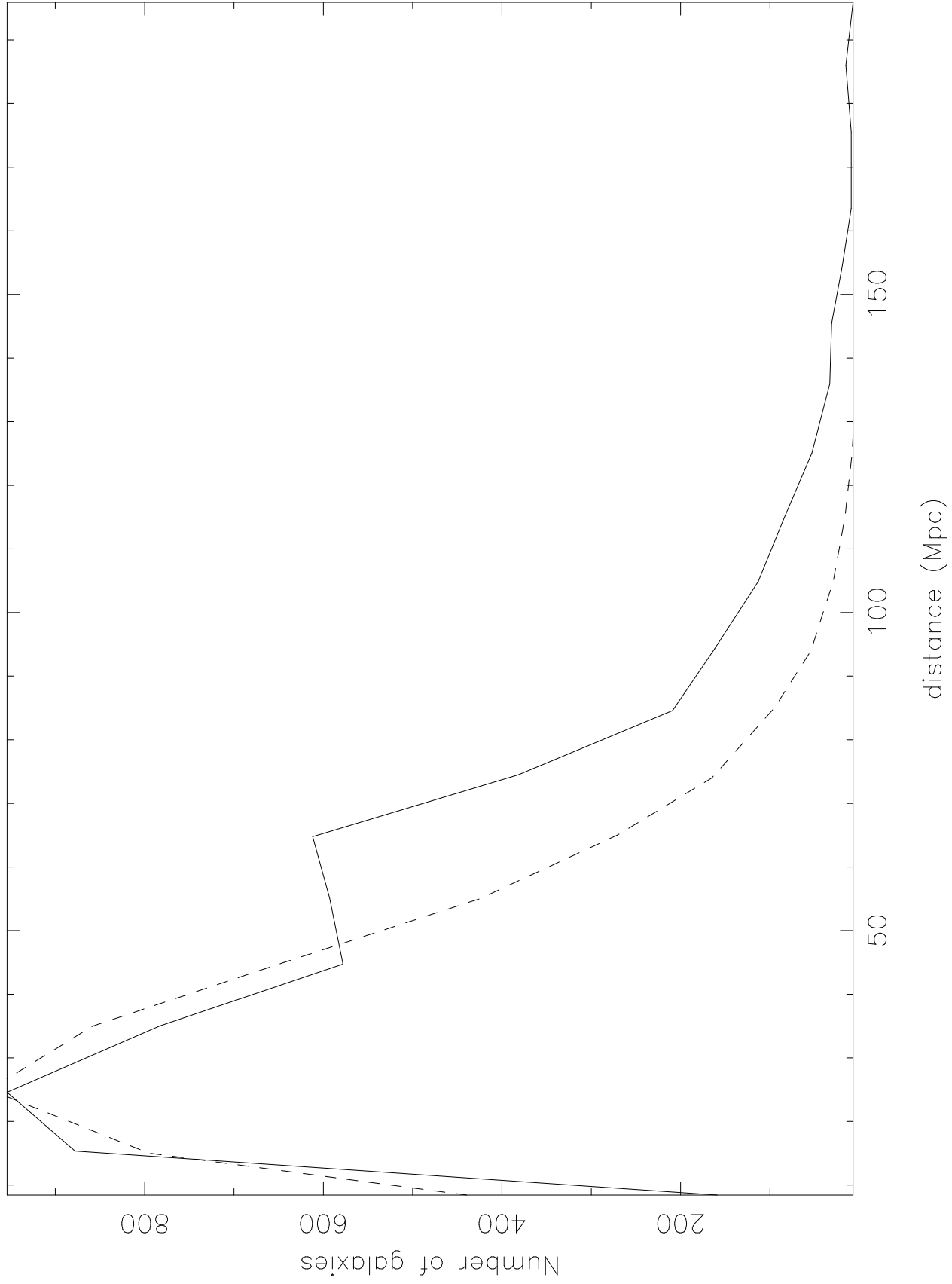
\end{document}